\documentclass[a4paper,11pt]{article}
\pdfoutput=1 

\usepackage{jheppub} 

\usepackage[T1]{fontenc} 

\usepackage{amssymb}
 \usepackage{amsthm}

\def\tr{\mathop{\rm tr}\nolimits}

\usepackage[latin1,utf8]{inputenc}
\usepackage{amsfonts}

\usepackage{xcolor}

\title{\boldmath Essential Fierz identities for a fermionic field}

\author[a]{Roberto Dale,}
\author[b,1]{Alicia Herrero \note{Corresponding author.}}
\author[c]{and Juan Antonio Morales-Lladosa}

\affiliation[a]{
Departamento de Estad\'istica, Matem\'atica e Inform\'atica and Centre of Operations Research (CIO),
Universidad Miguel Hernandez, E-03202 Elche, Alicante, Spain
}

\affiliation[b]{Institut de Matemàtica Multidisciplinar and Departament de Matem\`atica 
Aplicada,\\
Universitat Polit\`ecnica de Val\`encia, E-46022 Val\`encia, Spain. 
}

\affiliation[c]{Departament d'Astronomia i Astrof\'{\i}sica, Universitat de
Val\`encia, E-46100 Burjassot, Val\`encia, Spain; and
Observatori Astron\`omic, Universitat de
Val\`encia, E-46980 Paterna, Val\`encia, Spain.
}

\emailAdd{rdale@umh.es}  
\emailAdd{aherrero@mat.upv.es}
\emailAdd{antonio.morales@uv.es.}

\abstract{
For a single fermionic field, an interpretation of the Fierz identities (which establish relations between the bilinear field observables) is given. They appear closely related to the algebraic class (regular or singular) of the spin 2-form $S$ associated to the spinor field. If $S \neq 0$, the Fierz identities follow from the 3+1 decomposition of the eigenvector equations for $S$ with respect to an inertial laboratory, which makes this interpretation suitable for fermionic particle physics 
models. When $S= 0$, the Fierz identities reduce to three constraints on the current densities associated with the spinor field, saying that they are orthogonal, equimodular, the vector current being timelike and the axial one being spacelike.}

\begin{document} 
\maketitle
\flushbottom



\section{Introduction }
\label{sec:intro}

A fermionic field is usually described by a four-component spinor $\Psi$,  $\Psi (x) \in {\mathbb C}^4$ at each space-time event  $x$. Given a basis of the space-time exterior algebra, say $\{{\Gamma}^{A}\}_{A=1}^{16}$, a set of sixteen bilinear forms, $\Psi^\dag \Gamma^A \Psi$, can be constructed from  $\Psi$, where $\Psi^\dag$ 
is the hermitian conjugate spinor of $\Psi$. These bilinear forms reveal physical properties of the field and behave in a specific tensorial manner under Lorentz transformations. Substituting the hermitian conjugate spinor by the Dirac adjoint $\bar\Psi=\Psi^\dag \gamma^0$, these bilinear forms are called bilinear Dirac covariants (or local electron observables), where $\gamma^0$  is the Dirac conjugation matrix (see, for instance, \cite{Sakurai}). Reciprocally, $\Psi$ can be obtained (up to a global phase factor)  from its bilinear concomitants. This is the {\em spinor reconstruction theorem} 
(see Refs. \cite{Kaempffer-1981,Takahashi-1982,Takahashi-1983a,Takahashi-1983b,Crawford-1985,Holland-1986,
Crawford-1990}). 

The 16 bilinear covariants are not generically independent \cite{Laporte-Uhlenbeck-1931a,Uhlenbeck-Laporte-1931b,deBroglie-1934,Pauli-1936,Kofink-1937,Takabayasi-1957}.  In fact, in every space-time event, the bilinear forms are algebraic quantities, in the sense that their definition does not depend on the dynamics (Dirac, Klein-Gordon equations) associated with the field. Moreover, the products of any two bilinear forms, when expressed in terms of linear combinations of all of them, satisfy certain algebraic relations, which are usually referred as {\em Fierz identities}. They are derived from the completeness relations that give the canonical basis of the exterior algebra in terms of the basis under consideration $\{{\Gamma}^{A}\}_{A=1}^{16}$ (see, for instance, \cite{Nishi-2005, Minogin-2011, Minogin-2014}). The Fierz identities are also known as Pauli-Fierz identities \cite{Zhelnorovich-2019}.

A similar type of relations involving two different spinors, in particular a field and its derivative \cite{Minogin-2012}, appears when the energy tensor of the field is analyzed \cite{Inglis-Jarvis-2016}, or 
in the formulation the Maxwell-Dirac equations based on the bilinear covariants \cite{Inglis-Jarvis-2014}. For a compendium of historical references on this topic, see the introduction in \cite{Minogin-2011} and \cite{Takabayasi-1957}, where a continuous media model governed by the Dirac equation is studied. 

Originally, in the case of the $\beta$-{\em decay},  the Fierz identitities involved four interacting fermionic fields \cite{Fermi-1934,Fierz-1937}. An exhaustive treatment of the {\em generalized} Fierz identities, that is, for a quadruple of fermionic fields in an arbitrary dimension ---with application beyond the Standard Model of particle physics--- is given in Appendix B of Ref. \cite{TOrtin-2004}. Applications of some generalized Fierz relations for fermion interaction processes, including numerical implementations by hand, are reported in Ref. \cite{Nieves-Pal-2004}.  

The geometry attached to a spin 2-form is tacitly used in particle physics theories \cite{Bilenky, Lesov}. In fact, the Hamiltonian constructed from four interacting fermions takes into account the null currents when the symmetry under parity is broken by the interaction.  The null currents of the spin 2-form associated with a charged lepton and its associated neutrino enter in the Hamiltonian by means of the operators $C_\pm \equiv \frac{1}{2}(I \pm \gamma^5$). A linear combination of products of two bilinear covariants with the form $\bar{\Phi}_1 \gamma^\mu C_\pm  \Phi_2$ is proportional to the sum or the difference of the timelike current $J^\mu$ and the spacelike current $K^\mu$. Therefore, the null leptonic currents, $l$ and $n$, enter in the Hamiltonian of the interaction.  These currents $l$ and $n$ are associated with the fundamental directions of a regular spin 2-form, in this case constructed from a pair of leptons (for instance, an electron, $\Psi_e$,  and its associated neutrino $\Psi_{\nu_e}$). The notation used is introduced in next section.

This paper is devoted to establish the {\em essential} set of relations between the bilinear covariants attached to a single spinor field $\Psi$. By `essential' we means non redundant, mathematically manageable, and able to retain the whole physical information of the Fierz identities. 
In \cite{Minogin-2011}, Minogin obtained a reduced set of 21 Fierz type relations, providing six different geometric representations for the 16 electron observables, depending on the chosen laboratory frame and the spinor field parameterization. The  21 relations in Ref.  \cite{Minogin-2011} are not exactly the same as the 21 relations deduced directly from the algebraic structure of a regular spin 2-form $S$ in this work. Here, both sets of 21 identities are compared, establishing that are linearly related between them and deduced from the sixteen {\em essential} relations given by the algebraic structure of $S$. Refs. \cite{Minogin-2011,Minogin-2014,Minogin-2012} are written in a clear 3-dimensional Euclidean notation and goes beyond earlier works.

A 4-dimensional Lorentzian representation of the Fierz identities is done in Refs. \cite{Crawford-1985, Holland-1986, Zhelnorovich-2019}, where an overcomplete set of Fierz type relations is reported.  This Lorentzian approach will be here revisited and reinterpreted in terms 
of the invariant algebraic elements of the spin 2-form $S$. 

At this point, we would like to add a historical comment. In Ref. \cite{Laporte-Uhlenbeck-1931a} (page 1396,  footnote 17), Uhlenberg and Laporte acknowledge a private communication by Rainich concerning some results and a ``rigorous proof of the fact that the Dirac equations possess only two algebraic quadratic invariants''. In Ref. \cite{Uhlenbeck-Laporte-1931b},  this idea was further developed. It seems that Rainich considered the possibility of describing the electron spin observables constraints on the basis of the algebraic classification of electromagnetic fields in Minkowski space-time. This work goes on this Rainich's pioneering idea \cite{Rainich-1925}.

The paper is distributed as follows. To begin with, in Sec. \ref{sec:2}  the  necessary terminology to read the paper and some previous results related to algebraic properties of a Lorentzian antisymmetric covariant two-tensor (2-form) and 
to the bilinear covariants associated to a spinor are summarized. In Sec. \ref{sec:3}, the eigenvector equations for the spin 2-form and its (star or Hodge) dual 2-form are decomposed with respect to an arbitrary observer. In Sec. \ref{sec:4}, the complete independent set of identities between the local electron observables is written in a covariant form and its  equivalence to the eigenvector equations for the spin 2-form is established.  In Sec. \ref{sec:5}, our result is compared with the set of relations presented in Ref.  \cite{Minogin-2011}, where the standard 3-dimensional Euclidean notation is used. In Sec. \ref{sec:6}, the algebraic classification of a spinor field is presented, laying stress on the role played by the character (regular or singular) of the spin 2-form and the  Fierz identities for this classification.  A discussion about the results concerns to Sec. \ref{sec:7}, containing our conclusions. 

%

\section{Terminology and Preliminaries}
\label{sec:2}

The main sign conventions and notation adopted in this paper are as follows: 
 
 (i) $g$ is the metric of the the Minkowski space-time, with signature $( -, +,  +, + )$. Let $\{e_{\mu}\}_{\mu=0}^3$ be a space-time basis,  and  $\{\theta^\mu\}_{\mu=0}^3$ its algebraic dual, 
$\theta^\mu(e_\nu) = \delta^{\mu}_{\nu}$, with $\delta^{\mu}_{\nu}$ the Kronecker delta. In biunivocal correspondence with $\{\theta^\mu\}_{\mu=0}^3$, there exist four $\gamma$-matrices,  $\{\gamma^\mu\}_{\mu=0}^3$,  
that satisfy the Clifford algebra anti-commutation relations:
\begin{equation}\label{algebra}
\gamma^\mu\gamma^\nu+\gamma^\nu\gamma^\mu= - 2g^{\mu\nu}I, 
\end{equation}
where $g^{\mu \nu} = g(\theta^\mu,  \theta^\nu)$ and $I$ is the $4 \times 4$ unit matrix.%
%
 \footnote{Notice that there is no limitation on the causal character of the vectors in the basis. Dirac electron theory was originally presented  in a $\gamma$ matrix representation related to an orthonormal basis  \cite{Dirac-1928}. Later, Derrick \cite{Derrick-1981} analyzed the Dirac equation in some unusual basis constituted by four metrically symmetric vectors (or symmetric frame), and he used their attached  $\gamma$-matrix representations. A frame $\{ e_A\}_{A=1}^4$ is said to be a symmetric frame (for the metric $g$) if $g_{AA}=g(e_A,e_A)=\mu$ and $g_{AB}=g(e_A,e_B)=\nu$, $A\neq B$. Derrick's work would be further developed due to Derrick basis are representative of seven symmetric causal classes of relativistic frames \cite{RS-1992}. Thus, starting from a Derrick's symmetric frame \cite{Derrick-1981,RS-1992}, one obtains four $\gamma$-matrices, which are metrically indistinguishable. From a causal point of view, the relativistic space-time frames (and coordinate systems) have been classified in 199 causal classes \cite{199,199b}. Then, in accordance with the Clifford anti-conmmutation relations given by Eq. (\ref{algebra}),  there exists 199 causal classes of $\gamma$-matrix representations.  Such an abundance of $\gamma$ matrix representations could be used to describe fermion processes in non inertial frames, in order to develop further the research presented in  Ref. \cite{Chapman-Leiter-1976}.  \label{foot-basis}}
%

 (ii) The bilinear covariants associated with a spinor $\Psi$ and its Dirac adjoint $\bar{\Psi}=\Psi^\dagger \gamma^0$ are defined by (cf. \cite{Minogin-2011,Lounesto}),
\begin{equation}\label{bilinears}
\begin{array}{l}
\Omega_1= \bar{\Psi}\Psi ,\\ [0.2cm]
J^\mu= \bar{\Psi}\gamma^\mu\Psi , \\ [0.2cm]
S^{\mu\nu}= \frac{\rm{i}}{2} \bar{\Psi}(\gamma^\mu\gamma^\nu - \gamma^\nu\gamma^\mu) \Psi ,\quad  \\ [0.2cm]
K^\mu= \bar{\Psi}\gamma^\mu\gamma^5\Psi ,\\ [0.2cm]
\Omega_2= {\rm i}\, \bar{\Psi}\gamma^5\Psi ,
\end{array}
\end{equation}
where $\rm{i} = \sqrt{-1}$, and  $\gamma^5= - \rm{i} \gamma^0\gamma^1\gamma^2\gamma^3 $. 
Here,  $\Psi$ and its adjoint $\Psi^\dagger$ 
are represented by a column and a row complex matrix, respectively.

The bilinear covariants  (\ref{bilinears}) are phase independent and represent physical observables. In an orthonormal inertial frame,
\footnote{Accordingly with Eq. (\ref{algebra}) and the chosen signature $(-, +, +, +)$, the Dirac representation for the $\gamma$-matrices associated with an orthonormal basis is taken as
\begin{equation}\label{Dirac-r} \nonumber
\gamma^0_D=\left( \begin{array}{cc}
I_2 & 0_2 \\ 0_2 & - I_2
\end{array}\right) \quad , \quad \vec{\gamma}_D=\left( \begin{array}{cc}
0_2 & \vec{\sigma} \\ -\vec{\sigma} & 0_2
\end{array}\right) \, , 
\end{equation}
where $I_2$ ($0_2$) is the $2\times 2$ unit (zero) matrix, and $\vec{\sigma} = (\sigma_1, \sigma_2, \sigma_3)$ stands for the Pauli matrices,%
$$
\sigma_1=\left(\begin{array}{cc}
 0 & 1 \\
 1 & 0 
 \end{array}\right), \quad
\sigma_2=\left(\begin{array}{cr}
 0 & -i \\
 i & 0 
 \end{array}\right), \quad
\sigma_3=\left(\begin{array}{cr}
 1 & 0 \\
 0 & -1 
 \end{array}\right). \quad
$$
\label{Dirac-Pauli}}
they are  interpreted as density quantities that transform in a specific tensorial way under the action of the Lorentz group. They are called: the scalar $\Omega_1$,  the vector current $J=J_\mu\theta^\mu$,  the spin 2-form $S= S_{\mu\nu}\theta^\mu \otimes \theta^\nu = \frac{1}{2}S_{\mu\nu}\theta^\mu \wedge \theta^\nu$, the axial current $K=K_\mu\theta^\mu$ and the pseudoscalar $\Omega_2$, densities.  The metric $g$ is used to lower and rise indices. 

(iii) $\eta$ is the metric volume element of $g$, defined by
$\eta_{\alpha \beta \gamma \delta} = - \sqrt{- {\rm det}\, g}\, 
\epsilon_{\alpha \beta \gamma \delta}$,
where $\epsilon_{\alpha \beta \gamma \delta}$ stands for the  Levi-Civita 
permutation symbol, $\epsilon_{0 1 2 3} = 1$. 
The Hodge (or star) dual operator associated with $\eta$ is denoted by an asterisk $*$. For instance, the dual spin 2-form $*S$ has components 
$*(S)_{\mu\nu} = \frac{1}{2} \eta_{\mu \nu \lambda \rho} S^{\lambda \rho}$, and if $x, y, z$ are space-time 
vectors, one has that
\begin{equation}
[*(x \wedge y \wedge z)]_\alpha = \eta_{\alpha\beta\gamma\delta} x^\beta y^\gamma z^\delta, 
\end{equation}
where $\wedge$ stands for the {\em wedge or exterior product} (antisymmetrized tensorial product of totally antisymmetric tensors).

(iv)  Given $P$ and $Q$ second order tensors, the tensor $P \times Q$ denotes its matrix product, or contraction of adjacent indices, that is 
$$(P \times Q)_{\mu}\, ^{\nu} = P_{\mu \rho} \, Q^{\rho\nu}.$$ 
The trace of $P$ is $\tr (P) \equiv P^\mu_\mu$.
Then, for the spin 2-form $S$ one has 
\begin{equation} \label{trSS}
\tr(S^2) = \tr (S \times S)  = - S_{\mu\nu} S^{\mu\nu} ,
\end{equation}
\begin{equation} \label{trSdualS}
\tr (S \times * S) = - S_{\mu\nu} (*S)^{\mu\nu}.
\end{equation}
 
(v) For an observer of unit velocity $u$, $u^2\equiv g(u, u) = -1$,  any vector $x$ splits as: 
\begin{equation}
x = x^0 u +  \vec{x} = (x^0,\vec{x}),
\end{equation}
where $x^0 = - x \cdot u \equiv  - g(x, u)$  and  $\vec{x} \in E_\perp$ 
are the time-like and space-like components of $x$ relative to $u$, respectively, and
$E_\perp$ denotes the three-space orthogonal to $u$. 
Given two vectors $\vec{x}$, $\vec{y} \in E_\perp$, their vector or cross product is
expressed as
\begin{equation}\label{crossp}
\vec{x} \times\vec{y} = *(u \wedge \vec{x} \wedge \vec{y}).
\end{equation}

The interior or contracted product by $u$ is denoted by $i(u)$. For instance, if $S$ is a covariant $2$-tensor, one has that $[i(u)S]_\nu = u^{\mu} S_{\mu\nu}$ is a covector but if $S$ is a mixed $2$-tensor, then $[i(u)S]^\nu = u^{\mu} S_{\mu}^{\;\,\nu}$ is a vector. 

A 2-form $S$ is decomposed as
\begin{equation}\label{eh}
S=u \wedge e-*(u \wedge h),
\end{equation}
where $e=-i(u)S$ and $h=-i(u)*S$ are the electric and magnetic part of $S$ with respect to $u$, respectively.
Then,%
\footnote{For any space-time 2-form $F$, $*(*F) = - F$ . An extensive treatment on the algebraic properties of a space-time 2-form can be seen in Ref. \cite{Fradkin-1978,Tolo-Joan-1988}. 
In Ref. \cite{Tolo-Joan-1988}, the covariant determination of the eigendirections of $F$ is applied to characterize the differential conditions allowing the permanence of the null character of Maxwell fields.
\label{S-dual1}}
\begin{equation}\label{dual-eh}
* S= u \wedge h + *(u \wedge e).
\end{equation}

Note that to change $S$ by $*S$, $S \hookrightarrow *S$,  is equivalent to change the electric and magnetic parts of $S$ as $(e,h) \hookrightarrow (h, -e)$. %

(vi) The characteristic equation of a space-time 2-form $S$ is 
\begin{equation}\label{S-cha}
\lambda^4 -A \lambda^2 - \frac{B^2}{4}  = 0,
\end{equation}
where $A$ and $B$ are quadratic algebraic scalars defined as:
\begin{equation}\label{AB}
A \equiv \frac12 \tr (S^2) = e^2-h^2,\quad  B \equiv \frac12 \tr (S \times *S)=2\ e\cdot h.
\end{equation}
From (\ref{S-cha}), the eigenvalues of $S$ are $\pm \alpha$ and $\pm {\rm i} \beta$, with
\begin{equation}\label{alpha}
|\alpha| = {1 \over \sqrt{2}} \sqrt{\sqrt{A^2+B^2} + A} 
\end{equation}
and 
\begin{equation}\label{beta}
|\beta| = {1 \over \sqrt{2}}
\sqrt{\sqrt{A^2+B^2} - A} \ .
\end{equation}
Consequently, 
\begin{equation}\label{AB-bis}
A = \alpha^2-\beta^2,\quad B = 2 \alpha \beta \, .
\end{equation}

 (vii) A  2-form $S$ is said to be {\it regular} if $A^2 + B^2 > 0$. Otherwise, $A = B = 0$, and $S$ is said to be {\it singular}.  In terms of the eigenstructure of $S$ one has the following characterization:

(a) A  2-form $S$ is regular if, and only if, there exist two vectors $l, n$ and two algebraic invariants $\alpha, \beta$,  with $\alpha^2+\beta^2 \not=0$ such that%
\footnote{Vectors and covectors that are metrically equivalent are denoted with the same symbol without chance of confusion. For instance, $l_\mu=g_{\mu\nu} l^\nu$ is the covector associated to the vector $l$ by the metric. So, in the equation $i(l)S = - \alpha l$, the $l$ of the left side of the equation is a vector while the $l$ of the right side is a covector, when we consider $S$ as an antisymmetric covariant $2$-tensor. \label{metrequi}}
\begin{equation}\label{Sregular1}
\begin{array}{cl}
 i(l)S = - \alpha l\ , & i(n)S = \alpha n \ ,\\ [0.3cm]
 i(l)*S = \beta l\ , & i(n)*S = - \beta n.
\end{array}
\end{equation}
Then $l$ and $n$ are necessary null (light-like) vectors, $l^2 = n^2 =0$, and are called the {\em principal directions} of $S$. 
Moreover, $S$ and $*S$ admit the following canonical expressions
\begin{equation}\label{Sregular2}
S = \alpha \, n \wedge l +\beta *(n \wedge l), 
\end{equation}
\begin{equation}\label{dual-Sregular2}
*S = \beta \, l \wedge n - \alpha *(l \wedge n), 
\end{equation}
where, without loss of generality, we have chosen a parameterization of $l$ and $n$ such that $l \cdot n = -1$. 

(b) A  2-form $S$ is singular if, and only if,  there exists a vector $ l $ such that
\begin{equation}\label{Ssingular1}
i(l)S =i(l)*S =0. 
\end{equation}
Then $l$ is necessarily  null, $l^2 = 0$, and defines the {\em fundamental direction} of $S$, which can be expressed as
\begin{equation}\label{Ssingular2}
S = l \wedge p, 
\end{equation}
where $p$ is a determined space-like vector, up to the transformation  $\ p \hookrightarrow p+ \mu l$.
%
%


\section{Spin eigenvector equations: relative formulation}
\label{sec:3}
In this section we express the eigenvector equations of a spin 2-form $S$ with respect to a given inertial observer $u$.
For a regular $S$, Eqs. (\ref{Sregular1}) give the eigenstructure of $S$ and $*S$, where $\alpha$ and $\beta$ satisfy $\alpha^2 + \beta^2 > 0$ and $l$ and $n$ are the principal directions of $S$. 
For the observer $u$,  $l$ and $n$ decompose as
\begin{equation}\label{eleeneu}
l=l^0 u+\vec{l}, \quad n=n^0 u+\vec{n}\ .
\end{equation}
Then, the relation $i(l)S=-\alpha l$ is equivalent to:
\begin{equation}\label{eleS}
\alpha l^0 = \vec{l}\cdot \vec{e} \quad \textrm{and} \quad \alpha \vec{l} = l^0 \vec{e}+\vec{l}\times \vec{h} \, ,
\end{equation}
and the relation $i(n)S = \alpha n$ can be written as 
\begin{equation}\label{eneS}
\alpha n^0 = - \vec{n}\cdot \vec{e} \quad \textrm{and}\quad
\alpha \vec{n} = - n^0 \vec{e} - \vec{n}\times \vec{h} \, ,
\end{equation}
$\vec{e}$ and $\vec{h}$ being the electric and magnetic part of $S$ with respect to $u$, respectively.

Replacing $\vec{e}$ by $\vec{h}$ and $\vec{h}$ by $-\vec{e}$ by means of the Hodge duality,  
the relations  $i(l)*S = \beta l$, $i(n)*S = - \beta n$ in (\ref{Sregular1}) lead to
\begin{equation}\label{eleS-dual}
\beta l^0 = \vec{l} \cdot \vec{h} \quad \textrm{and} \quad \beta \vec{l} = l^0 \vec{h} - \vec{l}\times \vec{e} \, ,
\end{equation}
\begin{equation}\label{eneS-dual}
\beta  n^0 = - \vec{n} \cdot \vec{h} \quad \textrm{and}\quad \beta \vec{n} = - n^0 \vec{h} + \vec{n}\times \vec{e}, 
\end{equation}
respectively. Thus, the principal directions, $l$ and $n$, of a regular spin 2-form $S$ satisfy Eqs. (\ref{eleS})-(\ref{eneS-dual}).

For a singular $S$, the relations given in (\ref{Ssingular1}) lead to
\begin{equation}\label{eleS-singular}
\vec{l}\cdot \vec{e} = 0 \quad \textrm{and} \quad l^0 \vec{e}+\vec{l}\times \vec{h} = 0 \, ,
\end{equation}
\begin{equation}\label{eleS-dual-singular}
\vec{l} \cdot \vec{h} = 0 \quad \textrm{and} \quad l^0 \vec{h} - \vec{l}\times \vec{e} = 0 \, ,
\end{equation}
which can also be obtained just making $\alpha = \beta = 0$ in (\ref{eleS}) and  (\ref{eleS-dual}).
Thus,  the fundamental direction $l$ of a spin 2-form $S$ satisfies  Eqs. (\ref{eleS-singular})-(\ref{eleS-dual-singular}).

Moreover, the eigenvalues, $\pm\alpha$ and $\pm\rm{i}\,\beta$, of the spin 2-form $S$ have a direct relation with the bilinears $\Omega_1$ and $\Omega_2$ defined in (\ref{bilinears}). In fact, from the definition of $S$ in (\ref{bilinears}) one gets that
\footnote{To obtain these results we use the fact that the matrix $Z= 4\Psi\bar\Psi$ can be decomposed as
$$
Z=\Omega_1I_4+J_\mu\gamma^\mu+\frac{\rm{i}}{2}S_{\mu\nu}\gamma^\mu\gamma^\nu+K_\mu\gamma^\mu\gamma^5-\rm{i}\ \Omega_2\gamma^5,
$$ since $Z$ is not a general multivector but comes from the spinor $\Psi$.
\label{S-dual2}}
$$
\tr(S^2) =  2 (\Omega_2^2 - \Omega_1^2), \quad \tr (S \times * S)  = 4 \Omega_1 \Omega_2 .
$$
Then, from Eqs. (\ref{AB})-(\ref{AB-bis}), one realizes that
\begin{equation}\label{alphabeta}
\alpha =  \Omega_2 \quad {\rm and} \quad  \beta = \Omega_1.
\end{equation}
These relations (\ref{alphabeta}) are consistent with the common apellation used for $\Omega_1$ and $\Omega_2$  as scalar and pseudo-scalar quantities, respectively. Let $\{ e_\mu\}_{\mu=0}^3$ an orthonormal tetrad adapted to the geometry of a regular 2-form $S$, that is,
\begin{equation}\label{adapta-l-n}
l=\frac{1}{\sqrt{2}}(e_0+e_1), \qquad n=\frac{|}{\sqrt{2}}(e_0-e_1).
\end{equation}
The parity transformation $(e_0, e_1) \hookrightarrow (e_0, - e_1)$ changes $l$ by $n$ and the space-time volume element changes its sign, $\eta\hookrightarrow -\eta$. Then, $(S, *S) \hookrightarrow (S, - *S)$ and, according to Eqs. (\ref{Sregular2}) and (\ref{dual-Sregular2}), $(\alpha, \beta) \hookrightarrow (-\alpha, \beta)$, that is,
\begin{equation}\label{Canvi-Omega}
(\Omega_1, \Omega_2) \hookrightarrow (\Omega_1, - \Omega_2)\, ,
\end{equation}
under the considered spatial reflection.  

In the next section the algebraic structure of the spin 2-form $S$ is connected to the bilinear covariants associated with the fermionic field $\Psi$ and the Fierz identities they satisfy.


\section{Algebraic interpretation of the Fierz identities}
\label{sec:4}

In Ref. \cite{Crawford-1985}, an overcomplete set of Fierz identities between the bilinear covariants for any $S\neq0$ is reported. Using the notation introduced in Sec \ref{sec:2}, these identities are written as:%
\footnote{They are referred as Eq. (1.3) in \cite{Crawford-1985}, and apply for any (regular or singular) spin $2$-form. A similar set of relations for these identities has been also considered in \cite{Holland-1986}, and \cite{Zhelnorovich-2019} (pages 136, 137).\label{foot-Crawford}}
 %
\begin{equation}\label{Craw1}
i(J) S = - \alpha K,  \quad i(K) S =  - \alpha J,
\end{equation}
\begin{equation}\label{Craw2}
i(J) * S =  \beta K,  \quad i(K) * S =  \beta J,
\end{equation}
\begin{equation}\label{Craw3}
\alpha S  - \beta * S =  J \wedge K,  \quad  \beta S  + \alpha * S =  * (J \wedge K),
\end{equation}
\begin{equation}\label{Craw4}
\tr S^2 = 2 (\alpha^2 - \beta^2), \quad \tr (S \times *S) =  4 \alpha \beta .
\end{equation}

These overcomplete set can be reduced to an equivalent set of independent identities according with the following statements.

\begin{enumerate}
\item [(a)] {\em For a regular spin 2-form, $S \neq 0$, the Fierz identities for the corresponding fermionic field are the eigenvector equations for  $S$ and $*S$ given by (\ref{Sregular1}); they are a set of 16 one-component relations. Then, the spin 2-form can be written as
\begin{equation} \label{S-regular-canonica}
S = \alpha \, n \wedge l +\beta *(n \wedge l) \, ,  
\end{equation}
where $l$ and $n$ are the principal null directions of $S$, and $\alpha$ and $\beta$ provide the eigenvalues of $S$ and $*S$. }%
\item [(b)]{\em For a singular spin 2-form, $S \neq 0$, the Fierz identities for a fermionic field are the eigenvector equations for $S$ and  $*S$,  $i(l) S = i(l) *S = 0$; they are a set of 
8 one-component relations. Then, the spin 2-form is given by
\begin{equation}\label{S-singular}
S = l \wedge p \, , 
\end{equation}
where $l$ is the {\em fundamental direction} of $S$ and $p$ a space-like vector orthogonal to $l$.
}
\item [(c)] {\em For a zero spin 2-form, $S = 0$,  the Fierz identities 
for a fermionic field are equivalent to the existence of two non-collinear density spinor currents, $l$ and $n$, that are null, $l^2 = n^2 = 0$, and may be parametrized by taking $l \cdot n = -1$}. 
\end{enumerate}

Note that Eq. (\ref{Sregular1}) exclusively involves intrinsic algebraic elements associated with $S$: its eigenvalues and eigendirections. 
Moreover, expressions in (\ref{Sregular1}) can be applied for any spin 2-form $S \neq 0$, that is, when $S$ is regular or singular. 
The singular case corresponds to take $\alpha = \beta = 0$. 
On the other hand, when $S = 0$, the Fierz identities reduce to three algebraic scalar relations that constrain the causal character of the spinor current densities. %

In the following we present the proof of these statements in separated subsections and comment on some of their consequences.

\subsection{Regular spin 2-form}
\label{subsec:4a}

In order to justify the statement (a), let us consider a regular spin 2-form $S$. From the principal directions $l$ and $n$ and the invariants $\alpha$ and $\beta$ appearing in Eqs. (\ref{Sregular1}), let us define %
\begin{equation}\label{J}
J_\xi \equiv \sqrt{\frac{1}{2} (\alpha^2 + \beta^2)}  \, (e^{\xi} l + e^{-\xi}{n})
\end{equation}
and 
\begin{equation}\label{K}
K_\xi \equiv \sqrt{\frac{1}{2} (\alpha^2 + \beta^2)} \, (e^{\xi} l - e^{-\xi}{n})\ ,
\end{equation}
where $\xi$ is a real parameter. Since $l$ and $n$ are null vectors, $l^2 = n^2 = 0$, and taking  $l \cdot n = -1$ one has%
\begin{equation}\label{JK-productes}
- J_{\xi}^2 = K_{\xi}^2 = \alpha^2 + \beta^2 > 0, \qquad J_\xi \cdot K_\xi = 0,
\end{equation}
for all $\xi \in \mathbb{R}$. Moreover, a direct calculation leads to that for every real value $\xi$, the double one-parametric family of currents $\{(J_\xi, K_\xi)\}_{\xi \in \mathbb{R}}$ satisfies the relations (\ref{Craw1})-(\ref{Craw4}).  Therefore, a regular spin 2-form $S$ determines the spinor currents $J$ and $K$, up to a boost in the timelike plane $\{l, n\}$ expanded by the principal directions of  $S$, that is
\begin{equation}\label{boost-ln}
\begin{array}{l}
J_\xi   =  \cosh \xi \, J + \sinh \xi \, K \ ,\\ \\
K_\xi  =   \sinh \xi \, J + \cosh \xi \, K \ .
\end{array}
\end{equation}
For $\xi = 0$, the spinors currents $J_0 \equiv J$ and $K_0 \equiv K$ are the associated to the spinor $\Psi_0\equiv\Psi$ as given in (\ref{bilinears}); and the relations (\ref{JK-productes}) are called the 
bilinear Pauli identities (see Ref. \cite{Kaempffer-1981}, and references therein). In addition, the pairs $(J_\xi, K_\xi)$ satisfy  
Eqs. (\ref{JK-productes}) and are the currents for a family of spinors $\{\Psi_\xi\}_{\xi \in \mathbb{R}}$, which share the same spin 2-form $S$ for any value of $\xi$.

Note that, at each space-time event,  for an observer $e_0$ in the 2-plane $\{l, n\}$ and a unit space-like vector of this plane, $e_1$ , orthogonal to $e_0$, the parity transformation $(e_0, e_1) \hookrightarrow (e_0, - e_1)$ interchanges the null currents $l$ and $n$ each in another and tranforms the functions $(\alpha, \beta)$ into $(-\alpha, \beta)$. Consequently, from Eqs. (\ref{S-regular-canonica}) and (\ref{S-singular}), the family of currents $\{(J_\xi, K_\xi)\}_{\xi \in {\mathbb{R}}}$ remains invariant under the above parity transformation, but its individual currents transform as 
$J_\xi \hookrightarrow J_{-\xi}$ and  $K_\xi \hookrightarrow - K_{-\xi}$. 
On the other hand, from the time inversion $(e_0, e_1) \hookrightarrow (- e_0, e_1)$ one has 
$(l, n) \hookrightarrow (- n, -l)$, and then $(J_\xi, K_\xi) \hookrightarrow (- J_{-\xi}, K_{-\xi})$.

The spinor reconstruction theorem could be reformulated (for the regular case) in terms of a set of seven elements (the 6 quantities of $S$ and a real boost parameter $\xi$). Moreover, this reformulation could be expressed explicitly in terms of the eigenstructure of $S$ and the $\xi$ parameter. In fact, the eigenvalues of $S$ are explicitly obtained in terms of $S$ from (\ref{alpha}) and (\ref{beta}). Nevertheless, the explicit obtention of $l$ and $n$ in terms of $S$ requires to apply the projection method (used in \cite{Fradkin-1978, Tolo-Joan-1988}), which is based on the minimal polynomial equation that $S$ satisfies. 

\subsubsection{Relative formulation of the Fierz identities}
\label{subsubsec:4a1}
From now on, the index $\xi$ will be omitted in the spinor currents without loss of generality, since the relations (\ref{Craw1})-(\ref{Craw4}) are satisfied for any $\xi\in{\mathbb R}$, for the given regular spin 2-form $S$.

Note that 
the statement (a) also allows to express the Fierz identities in a 3-dimensional formulation. 
In order to do it, let us split $J$ and $K$ as $J = (j^0, \vec{j})$ and  $K = (k^0, \vec{k})$, for a given inertial observer $u$. Then, by addition and subtraction of the scalar equations (time-like parts) in Eqs. (\ref{eleS}), (\ref{eneS}), (\ref{eleS-dual}) and (\ref{eneS-dual}) between them, one gets:
\begin{eqnarray}
\alpha j^0 & = & \vec{k} \cdot \vec{e}\ , \label{alfake}\\
\alpha k^0 & = & \vec{j} \cdot \vec{e}\ ,\label{alfaje}\\
\beta j^0 & = & \vec{k} \cdot \vec{h}\ ,\label{betakh}\\
\beta k^0 & = & \vec{j} \cdot \vec{h} \ ,\label{betakjh}
\end{eqnarray}
where we have taken into account that  $\alpha^2 + \beta^2 > 0$. Similarly, for the space-like parts of the same Eqs. (\ref{eleS}), (\ref{eneS}), (\ref{eleS-dual}) and (\ref{eneS-dual}), one obtains
\begin{eqnarray}
\alpha \vec{j} & = & k^0 \vec{e} + \vec{k} \times \vec{h}\ , \label{alfajkekh}\\
\alpha \vec{k} & = & j^0 \vec{e} + \vec{j} \times \vec{h}\ , \label{alfakjejh}\\
\beta \vec{j} & = & k^0 \vec{h} - \vec{k} \times \vec{e}\ , \label{betajkhke}\\
\beta \vec{k} & = & j^0 \vec{h} - \vec{j} \times \vec{e}\ . \label{betakjhje} 
\end{eqnarray}

Then, Eqs. (\ref{alfake})-(\ref{betakjhje}) provide the local physical interpretation (i. e. the relative formulation with respect a local space-time observer) of the eigenvector equations for a regular spin 2-form.  In Sec. \ref{sec:5}, this result is compared with the one obtained in Ref. \cite{Minogin-2011}.
%

\subsection{Singular spin 2-form}
\label{subsec:4b}

In order to justify the statement (b), let us consider a singular spin 2-form $S$ and its fundamental direction $l$, which satisfies Eqs. (\ref{eleS-singular})-(\ref{eleS-dual-singular}). Now, $l$ is the unique fermionic null density current. Then, the singular case may be understood as a limit situation of the regular case when $\alpha^2+\beta^2\to 0$.  These can be interpreted in two different ways from expressions (\ref{JK-productes}):

(b1) The two fermionic currents are light-like and colinear. This is equivalent to consider that $\xi \to \pm \infty$  in expressions (\ref{boost-ln}) and corresponds to take a boost in the 2-plane $\{l, n\}$ whose velocity goes to the light velocity ($\tanh \xi  \to \pm 1$). 

(b2) One of the currents becomes ligth-like and the other one goes to zero, that is, $J_\xi \equiv l$ and $K_\xi =0$,  or reciprocally. In fact, Eq. (\ref{eleS-singular}) is equivalent to Eqs. (\ref{betakjh}) and (\ref{betakjhje}) by taking $\alpha = 0$, or $(k^0, \vec{k}) = 0$, and replacing  $(j^0, \vec{j}) \hookrightarrow (l^0, \vec{l})$. In a similar way,  Eq. (\ref{eleS-dual-singular}) is equivalent to Eqs. (\ref{betakjh}) and (\ref{betakjhje})  by taking $\beta = 0$ and the above replacement, $(j^0, \vec{j}) \hookrightarrow (l^0, \vec{l})$. 

Thus, for the singular case, only the 8 relations given in (\ref{eleS-singular})-(\ref{eleS-dual-singular}) are non-trivial and significantly encode the Fierz identities.

\subsection{Zero spin 2-form, $S = 0$}
\label{subsec:4c}

Statement (c) says that there exist two density equimodular and orthogonal four-currents, $J$ and $K$, one being time-like and the other one being space-like, that is $J^2 = - K^2 < 0$ and $J \cdot K = 0$. In fact, one can realize that the 21 relations obtained in Ref. \cite{Minogin-2011}  reduce to the three constraining relations on the vector and the axial density currents associated with the spinor field (see Eqs.  (\ref{corrents-quadrats}) and (\ref{jk-orto}) later, in the next section).
Actually, there exist a one-parameter family of spinors having $S=0$ and two currents (\ref{boost-ln}) defining the same timelike 2-plane.

%


\section{Fierz identities and electron local observables constraints}
\label{sec:5}

In Ref. \cite{Minogin-2011} a set of $21$ algebraic equations relating the $16$ electron local observables (there denoted as $j_0$, $\vec{j}$, $f_0$, $\vec{f}$, $b$, $g$, 
$\vec{c}$ and $\vec{d}$)  is given. In this section, this set of equations is compared with the ones obtained in the previous section.  To make the comparison clearer, we first report the correspondence between our notation and the one used in \cite{Minogin-2011}:
\begin{eqnarray}
\alpha & \to & g\ , \\
\beta & \to & b\ , \\
\vec{e} & \to & - \vec{d}\ , \\
\vec{h} & \to & - \vec{c}\ , \\
(j^0, \vec{j})& \to & (j^0, \vec{j})\ ,\\
(k^0, \vec{k})& \to & (f^0, \vec{f})\ .
\end{eqnarray}

Next, the set of 21 relations established in Sec. \ref{sec:4} is rewritten using this correspondence in notation
and conveniently ordered as:

\begin{enumerate}
\item Currents constraining relations (3 equations): 
\begin{equation}\label{corrents-quadrats}
 j_0^2 - \vec{j}^{\, 2} = b^2 + g^2 = - f_0^2 + \vec{f}^{\, 2}, 
\end{equation}
\begin{equation}\label{jk-orto}
j_0 f_0 = \vec{j} \cdot \vec{f}, 
\end{equation}
which correspond to Eqs. (\ref{JK-productes}) since the scalar products are referred to an inertial frame, as well as $j_0 = -j^0$ and $f_0 = -f^0$, according with the chosen Minkowski metric signature. These three equations correspond to Eq. \underline{(31)} in Ref. \cite{Minogin-2011} (here underlined).

\item The spin algebraic relations derived from $S$ (2 equations):
\begin{eqnarray}
b g & = &  \vec{c} \cdot \vec{d} \label{S-invariant-1}\,\\
b^2 - g^2 & = & \vec{c}^{\, 2} - \vec{d}^{\, 2}\label{S-invariant-2},
\end{eqnarray}
which are Eqs. (\ref{AB}) and (\ref{AB-bis}). The first equation,
Eq. (\ref{S-invariant-1}), corresponds to Eq. \underline{(26.2b)},  and the second one, Eq. (\ref{S-invariant-2}), results by subtracting  \underline{(27.3)} from 
\underline{(27.2)} in Ref. \cite{Minogin-2011}.

\item The scalar relations among the 16 bilinear forms (4 equations)
\begin{eqnarray}
g j^0 & = & - \vec{f} \cdot \vec{d} \label{escalar-1}\,\\
g  f^0 & = & - \vec{j} \cdot \vec{d} \label{escalar-2}\,\\
b j^0 & = & - \vec{f} \cdot \vec{c} \label{escalar-3}\,\\
b f^0 & = & - \vec{j} \cdot \vec{c}\label{escalar-4}\, 
\end{eqnarray}
that is, Eqs. (\ref{alfake}), (\ref{alfaje}), (\ref{betakh}) and (\ref{betakjh}), which,
in increasing numbered order, correspond to Eqs. \underline{(26.3b)}, \underline{(26.1b)}, \underline{(26.3a)}
\underline{(26.1a)} in \cite{Minogin-2011}, respectively.

\item The 3-vectorial Euclidian relations among the 16 bilinear forms (12 
equations)
\begin{eqnarray}
g \vec{j} & = & - f^0 \vec{d}  - \vec{f} \times \vec{c} \label{vector-1}\ ,\\
g \vec{f} & = & - j^0 \vec{d} - \vec{j} \times \vec{c} \label{vector-2}\ ,\\
b \vec{j} & = & - f^0 \vec{c} + \vec{f} \times \vec{d}\label{vector-3}\ ,\\
b \vec{f} & = & - j^0 \vec{c} + \vec{j} \times \vec{d}\label{vector-4}\ , 
\end{eqnarray}
that is, Eqs. (\ref{alfajkekh}), (\ref{alfakjejh}), (\ref{betajkhke}) and (\ref{betakjhje}), which,
 in increasing numbered order, correspond to Eqs.  \underline{(28.6)}, \underline{(28.2)}, \underline{(28.7)}, \underline{(28.3)} in \cite{Minogin-2011}, respectively.
\end{enumerate}

Therefore, we have shown that the set of 21 equations given in \cite{Minogin-2011} is equivalent to the specification of the algebraic structure of $S$ and $*S$. Furthermore, up to linear combinations, this set transforms into Eqs. \underline{(36)}, \underline{(37)} and \underline{(38)} in Ref. \cite{Minogin-2011}, which are, respectively,  
6 escalar type equations, 9 vector type equations and 6 tensor type one component relations. 
%
%


\section{The role of the spin 2-form in the spinor classification}
\label{sec:6}

The algebraic classification of a four component space-time spinor (Lounesto classification) \cite{Lounesto}
is made by means of its bilinear covariants, given by (\ref{bilinears}), taking into account the Fierz identities. This classification splits in six algebraic types of spinors, which are characterized according to the possible nullities of the algebraic invariants $\alpha$ and $\beta$, the spin 2-form $S$, and the current $K$ (the current $J$ is taken nonzero) \cite{Lounesto}:
\begin{enumerate}
\item [(i)] $\alpha \neq 0, \beta \neq 0$.
\item [(ii)] $\alpha = 0, \beta \neq  0$.
\item [(iii)] $\alpha\neq 0, \beta = 0$.
\item [(iv)] $\alpha = \beta = 0, S \neq 0, K \neq 0$.
\item [(v)] $\alpha = \beta = 0, S \neq 0, K = 0$.
\item [(vi)] $\alpha = \beta = 0, S = 0, K \neq 0$.
\end{enumerate}

The Lounesto classification of a spinor points out the role played by the spin 2-form $S$ in distinguishing the general class to which a fermionic field belongs. From an algebraic point of view, the generic class $S\neq 0$ splits into two subclasses with different  algebraic character (regular or singular) of $S$. 

The regular class splits in the types (i), (ii) and (iii) of the Lounesto classification, and corresponds to Dirac fermions fields while the singular class splits in the type (iv) of flag dipole spinors and the type (v) of flag pole-spinors. Concretely, types (iv) and (v) are
identified with cases (b1) and (b2) obtained as a limit situation of the regular case in Subsection \ref{subsec:4b}, respectively.%
\footnote{This two types have been further analyzed in \cite{Cavalcanti-2014}. Moreover, type (iv) has also been recently studied 
 in  Refs. \cite{CoVillalobos-HdaSilva-daRocha-2015, CoVillalobosBAB-2020} where the possibility $J = 0$ is also contemplated, and type (v) contains the Majorana fermions and  the Elko fermions, both being eigenspinors of the charge conjugation operator (see Refs. \cite{Ahluwalia-2019,Ah-Gru-2005-PRD,Ah-Gru-2005,Ro-Ro-2006,daRocha-HdaSilva-2007,HdaSilva-daRocha-2013,HdaSilva-CoVillalobos-BRS-2016}).\label{foot-Lounesto}}

The non generic class $S = 0$ is the Lounesto type (vi) and contains the Weyl (or massless) neutrino field.%

The null eigendirections of $S$ and $*S$ does not change under a rotation of dualitity,
\begin{equation}\label{rotadualitat}
\begin{array}{rl}
\tilde{S} &= (\cos \theta) S  + (\sin \theta) *S,\\[0.2cm] 
* \tilde{S} &= - (\sin \theta) S + (\cos \theta) *S,
\end{array}
\end{equation}
but $\alpha$ and $\beta$ transform according with:
\begin{equation}\label{rd-alpha-beta}
\tilde{\alpha} = \alpha \cos \theta - \beta  \sin \theta ,\quad  \tilde{\beta} = \alpha \sin \theta + \beta \cos \theta ,
\end{equation}
and keeping that $\tilde{\alpha}^2 + \tilde{\beta}^2 = \alpha^2 + \beta^2$. Consequently, the current family $\{J_\xi,K_\xi\}_{\xi \in \mathbb{R}}$ given by (\ref{J}) and (\ref{K}) is invariant under the transformation (\ref{rotadualitat}). But, the quadratic scalars $A$ and $B$, given by (\ref{AB}), transform as a rotation of angle $2\theta$,
\begin{equation}\label{rd-AB}
\tilde{A} = A \cos 2\theta - B  \sin 2\theta ,\quad  \tilde{B} = A \sin 2\theta + B \cos 2\theta \ ,
\end{equation}
or equivalently,
\begin{equation}\label{rd-AB2}
\begin{array}{rcl}
\tilde{\alpha}^2-\tilde{\beta}^2 &= &(\alpha^2-\beta^2) \cos 2\theta - 2\alpha\beta  \sin 2\theta ,\\[0.2cm]
2\tilde{\alpha}\tilde{\beta}&=& (\alpha^2-\beta^2) \sin 2\theta + 2\alpha\beta  \cos 2\theta  .
\end{array}
\end{equation}

For this reason, one has that a duality rotation does not change the regular or singular character of the 2-form $S$ but it can move the spinor from one spinor type to another one inside the same subclass (regular or singular) in the spinor classification.


\section{Conclusions}
\label{sec:7}
In particle physics, for example in the $\beta$-decay theory or similar situations in electroweak interactions, Fierz identities involve algebraic constraints between interacting fermionic fields. To deepen in the essential information that such identities encode, we have considered the case of a single fermionic field. Then, expressing the eigendirections of a Lorentzian 2-form with respect to an inertial observer (Sec. \ref{sec:3}), the Fierz identities are interpreted in terms of the eigenstructure of the spin density tensor $S$, both in the regular and the singular cases (Sec. \ref{sec:4}).  This study displays how large  Fierz identities appear closely related to the space-time algebra.

In Ref. \cite{Minogin-2011} by Minogin, a set of 21 identities that the bilinear electron observables satisfy were obtained and written in a suitable 3-dimensional Euclidean notation (see Sec. \ref{sec:5}, which summarizes Sections 4 and 5 of Ref. \cite{Minogin-2011}). In this paper, a geometric interpretation of an equivalent set of identities based on the eigenvector relations fulfilled by $S$ and $*S$ has been performed. The differences between both sets have been analyzed. 
In Section \ref{sec:6}, the relation of the Fierz identities and the duality rotation with the algebraic types of the spinor classification has been established.

The algebraic interpretation of the essential Fierz identities between the bilinear covariants constructed with two spinors in terms of their spin 2-forms requires further investigation. 
The existence or not of any connection between the eigenstructures of a pair of interacting spin 2-forms should be the starting point to extend forward the present research: (i) the interpretation of the Fierz identities for a pair of fermion fields, and (ii) the algebraic classification of the energy tensor of a sole fermionic field  \cite{Inglis-Jarvis-2016}, which involves the bilinear covariants constructed from the field and its first derivatives. 

Originally, the Fierz identities come from the restrictions that a change of basis in the space-time induces on the basis of the Minkowski exterior algebra (specified by four vectors, six 2-planes and four 3-planes). This geometric vision of the whole Fierz identities is related with the causal classification of the the space-time frames \cite{199}. In fact, this classification results from the analysis of the whole set of  constraints between the causal characters of the 14 geometric elements of a space-time frame: its 4 directions, its 6 two-planes and its 4 three-planes. This consideration will help to explore whether there exists a deeper connection between the Lorentzian structure of the space-time geometry and the fundamental interactions between elementary particles.   
%


\acknowledgments

This work has been supported by the Spanish Ministerio de Ciencia, Innovaci\'on y Universidades, Projects PID2019-109753GB-C21 and PID2019-109753GB-C22.


\end{document}